\documentstyle[epsfig]{aipproc}

\def\frac#1#2{ {{#1} \over {#2} }}

\def\beq{\begin{displaymath}}
\def\eeq{\end{displaymath}}

\def\Q_s{\mu}

\newcommand{\lwig}{\mbox{\,\raisebox{.3ex}
    {$<$}$\!\!\!\!\!$\raisebox{-.9ex}{$\sim$}\,}}
\newcommand{\gwig}{\mbox{\,\raisebox{.3ex}
    {$>$}$\!\!\!\!\!$\raisebox{-.9ex}{$\sim$}}\,}

\newcommand{\ii}{{\rm i}}

\def\np#1#2#3{Nucl.\ Phys.\ {\bf B#1}, #2 (19#3)}
\def\pl#1#2#3{Phys.\ Lett.\ {\bf #1B}, #2 (19#3)}
\def\pr#1#2#3{Phys.\ Rev.\ D {\bf #1}, #2 (19#3)}

\begin{document}
\title{QCD -- Instantons in $e^\pm P$ Scattering\thanks{Talk presented at 
    the 5th
    International Workshop on Deep-Inelastic Scattering and QCD (DIS\,97),
    Chicago, April 1997; to be published in the Proceedings.}}

\author{S. Moch, A. Ringwald and F. Schrempp}
\address{DESY, Notkestr. 85, D-22603 Hamburg, Germany}


\begin{flushright}
DESY 97-114 \\
hep-ph/9706400
\end{flushright}

\maketitle

\begin{abstract}
We review recent theoretical results from our systematic study of QCD-instanton
contributions to $e^\pm P$ processes, involving a hard momentum scale 
$\mathcal{Q}$.  The main issues are: the absence of IR divergencies due
to a dynamical suppression of instantons with large size 
$\rho > 1/\mathcal{Q}$, the reliable calculability of instanton-induced 
amplitudes and our inclusive framework to 
systematically calculate properties of the $I$-induced multi-parton 
final state.
\end{abstract}

\section*{Introduction}

Instantons~\cite{bpst} are well known to represent topology changing 
tunnelling transitions in non-abelian gauge theories. 
These transitions induce processes which are {\it forbidden} in perturbation
theory, but have to exist in general~\cite{th} due to Adler-Bell-Jackiw
anomalies. Correspondingly, these processes imply  a violation of 
certain fermionic quantum numbers, notably, $B+L$ in the electro-weak gauge
theory and chirality ($Q_5$) in (massless) QCD.

An experimental discovery of such a novel, non-perturbative
manifestation of non-abelian gauge theories would clearly be of basic
significance.

The interest in instanton ($I$)-induced processes
during recent years has been revived by the observation~\cite{r} 
that the strong exponential suppression, $\propto \exp (-4\pi
/\alpha)$,  of the corresponding tunnelling rates 
 {\it at low energies}  may be overcome
at {\it high energies}, mainly due to multi-gauge boson emission in
addition to the minimally required fermionic final state. Since 
$ \alpha_s \gg \alpha_W$, QCD-instantons are certainly much less
suppressed than electro-weak ones. A pioneering and encouraging
theoretical estimate of the size of the QCD-instanton  induced 
contribution to the gluon structure functions in deep-inelastic 
scattering (DIS)  was recently presented in Ref.~\cite{bb}.  

A systematic phenomenological and theoretical study
is under way~\cite{rs,grs,rs1,ggmrs,mrs,rs2,rs3}, which  clearly 
indicates that deep-inelastic $e^\pm P$ scattering at HERA now offers a 
unique window to experimentally detect QCD-instanton induced processes
through their characteristic {\it final-state signature}. 
While our phenomenological approach and the ongoing experimental searches for
QCD instantons were reviewed in WG\,III~\cite{ar,c}, we shall focus here
on our recent theoretical results. 

They clearly indicate that $e^\pm P$ scattering, involving a hard
momentum scale $\mathcal{Q}$, plays a distinguished r{\^o}le for 
studying manifestations of QCD-instantons. This mostly refers to 
the DIS regime, but possibly also to hard photoproduction, where 
$\mathcal{Q}$ denotes the large transverse momentum of a jet.

In this talk, we shall concentrate on the following important issues: 

We set up the relevant $I$-induced amplitudes for DIS at the parton
level in leading semi-classical approximation and outline why they are
well-defined and calculable for small $ \alpha_s({ \mathcal{Q}})$. 
We concentrate on the crucial feature
that the generic IR divergencies from integrating over the 
$I$-size $\rho$ are absent in DIS, since the hard momentum scale 
$ \mathcal{Q}$ provides a dynamical cutoff,
$ \rho\lwig{\mathcal O}(1/{ \mathcal Q})$. As an example, the cross
section for the simplest $I$-induced process is explicitly evaluated
and compared to the corresponding contribution from perturbative QCD.
Finally, we briefly sketch our inclusive framework to 
systematically calculate properties of the $I$-induced multi-parton 
final state. It accounts for the exponentiation 
of produced gluons including final-state tree-graph corrections.

\section*{2. Instanton-Induced Processes in Leading 
             Semi-Classical Approximation }
The main $I$-induced contribution to deep-inelastic $e^\pm P$
scattering  comes from  the processes,  
       $$\gamma^\ast + {\rm g} \Rightarrow 
       \sum_{\rm flavours}^{n_{f}}\left[\overline{{\rm q}_L}+{\rm q}_R\right] 
       + n_{\rm g}\,{\rm g},$$
which correspond to $\triangle Q_5 = 2\,n_f$ and thus
vanish to any order of conventional perturbation theory  
in (massless) QCD.

The evaluation of the corresponding amplitudes involves the following steps:

We start with the basic building blocks in Euclidean configuration
space (see e.g. Ref.~\cite{mrs}), the classical instanton gauge field  
$A_\mu ^{(I)}\,(x;\rho,U)$, the quark zero modes 
$\kappa^{( I)}\,(x;\rho,U),\  \overline{\phi}^{( I)}\,(x;\rho,U)$  
and the (non-zero mode) quark propagators in the $I$-background 
$S^{(I)}\,(x,y;\rho,U), \overline{S}^{(I)}\,(x,y;\rho,U)$.
The classical fields and quark propagators depend on collective
coordinates, the $I$-size $\rho$ and the colour orientation matrices 
$U$. 
   
``Instanton-perturbation theory'' is generated by expanding
the (Euclidean) path integral for the relevant Green's functions 
about the classical instanton solution for small $\alpha_s$. 
Next, Fourier-transforms (FT) to momentum space with respect 
to external lines are performed and the external legs are LSZ
amputated. Finally, the result is analytically continued to Minkowski space.

After performing the FT's with respect to the external lines,  the 
leading-order amplitude  in Euclidean space takes the
following form~\cite{mrs} for the simplified case of one flavour
($n_{f}=1$) (see Fig.\,1),
       
\begin{figure}[t!]
\vspace{-0.4cm}
\begin{center}
\epsfig{file=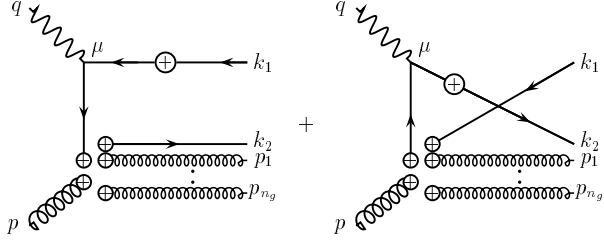,%
width=8cm}
\caption[dum]{$I$-induced contribution to DIS for $n_f=1$ in 
leading semi-classical approximation. }
\end{center}
\end{figure}

 \begin{eqnarray}
\lefteqn{
  {\mathcal T}_\mu ^{({ I})\,a\,a_1\ldots a_{n_g}} 
    =-\ii\, e_{q}\, { d\,\left( \frac{2\,\pi}{\alpha_s(\mu _{r} )}\right)^6\, 
     {\exp}\left[-\frac{2\,\pi}{\alpha_s(\mu _{r} )}\right]}  
  \int dU\,\int\limits_0^\infty
  \frac{d\rho}{\rho^5}\, 
    \left(
    \rho\,\mu _{r} \right)^{\beta_0}\, \times}
\nonumber
\\
&&  \lim_{p^{2}\rightarrow 0}p^{2}\,
    {\rm tr}\left[ \lambda^{a}\,
    \epsilon_g(p)\cdot A^{(I)}(p;\rho,U)\right]  
    \prod_{i=1}^{n_g}\,\lim_{p_i^{2}\rightarrow 0}p_i^{2}\,
    {\rm tr}\left[ \lambda^{a_i}\,\epsilon^{\ast}_g(p_i)\cdot 
    A^{(I)}(-p_i;\rho,U)\right]\,    
   \times 
\nonumber
\\  &&
   \chi_{R}^{\dagger}(k_{2})\, 
    \Biggl[ 
     \lim_{k_{2}^{2}\rightarrow 0}
     \,(\ii k_{2})\,\kappa^{(I)} (-k_{2};\rho,U)\,
     \lim_{k_{1}^{2}\rightarrow 0}\,
     { {\mathcal V}_\mu ( q,{ -k_{1}};\rho,U)  }\Biggr. 
\label{Tlsc}
\\  &&
   \Biggl. +   \lim_{k_{2}^{2}\rightarrow 0}\,
    { {\mathcal V}^c_\mu ( q, -k_{2};\rho,U)}\,
     \lim_{k_{1}^{2}\rightarrow 0}\,
    \overline{\phi}^{(I)}(-k_{1};\rho,U)\,
    (-\ii\,\overline{k}_{1}) 
     \Biggr] 
\chi_{L}(k_{1})\, ,
\nonumber
\end{eqnarray}
The instanton-density~\cite{th} with renormalization scale $\mu _{r}$
and 1st coefficient $\beta_0=11-2/3\, n_f$ of the perturbative 
QCD beta-function,
\begin{equation}
    d\,\left( \frac{2\,\pi}{\alpha_s(\mu _{r})}\right)^6\, 
     \exp\left[-\frac{2\,\pi}{\alpha_s(\mu _{r}) }\right] \,
     \left(\rho\,\mu _{r} \right)^{\beta_0}\, ;\ 
    d\ \mbox{\rm being a known constant,}
\end{equation}
is 1-loop renormalization group (RG) invariant.

First of all, note the strong IR divergence
(large $\rho$), if the $\rho$-integral in Eq.\,(\ref{Tlsc}) were 
performed with just the $I$-density. Hence, convergence may only 
come from  further $\rho$ dependence inherent in the matrix element. In 
particular, any possible  cut-off for large $\rho$ in terms of the inverse 
hard scale $1/\mathcal{Q}$, must be hidden in the FT'ed 
photon-fermion ``vertices'', ${\mathcal V}_\mu (q,-k_{1};\rho,U)$ 
and ${\mathcal V}^c_\mu ( q,-k_{2};\rho,U)$ (c.\,f. Figs.\,1,\,2). 
 
FT and LSZ-amputation of  the instanton 
gauge field $A^{(I)}_{\mu}$ and quark zero modes 
$\kappa$ and $\overline{\phi}$ is straightforward~\cite{r}.
They only contribute {\it positive} powers of $\rho$.
\begin{figure}[t!]
\vspace{-0.4cm}
\begin{center}
\epsfig{file=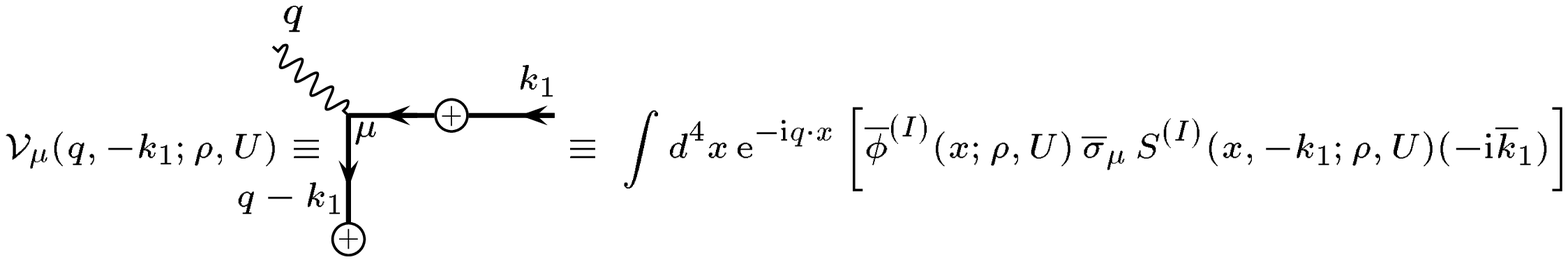,%
width=14cm}
\caption[dum]{The photon-fermion vertex, ${\mathcal V}_\mu (q,-k_{1};\rho,U)$,
generating the {\it dynamical cut-off} 
$\rho\lwig {\mathcal O}(1/{\mathcal Q})$.}
\end{center}
\end{figure}
On the other hand, the LSZ-amputation of the current quark in the
FT'ed photon-fermion vertices, ${\mathcal V}_\mu (q,-k_{1};\rho,U)$ 
and ${\mathcal V}^c_\mu ( q,-k_{2};\rho,U)$,  is quite non-trivial, since
they involve a two-fold FT of the complicated
quark propagators in the $I$-background (see Fig.\,2). 
After a long and tedious calculation, we find~\cite{mrs},
\begin{eqnarray}
\lim_{k_1^2\rightarrow 0}
{\mathcal V}_\mu
(q,-k_1;\rho,U)&=&2\pi\ii{ \rho^{3/2}}
\left[ \epsilon \sigma_\mu  
 \overline{V} (q,k_1;{ \rho }) 
\,U^\dagger \right] \,
\\
\lim_{k_2^2\rightarrow 0}
{\mathcal V}^c_\mu (q ,-k_2;\rho,U)&=&2\pi \ii { \rho^{3/2}}\,
\left[ U\,  V(q,k_2;{ \rho} ) 
\overline{\sigma}_\mu \epsilon \right] \, ,
\end{eqnarray}
where with the shorthand $q^\prime\equiv q-k$,
\begin{equation}
 V (q,k;{ \rho} ) =
\frac{k}{2 q\cdot k}{ \rho}\sqrt{q^2}\,
K_1\left({ \rho}\sqrt{q^2}\right)
+ \left[  
\frac{q^\prime}{q^{\prime\,2}}-\frac{k}{2 q\cdot k}
\right]
{ \rho}\sqrt{ q^{\prime\,2}}\,
K_1 \left({ \rho}\sqrt{q^{\prime\,2}}\right).
\label{vertex}
\end{equation}
Obviously, the integration over the instanton size $\rho$ in the 
amplitudes ${\mathcal T}^{(I)}_{\mu}$, Eq.\,(\ref{Tlsc}),  is finite due   
to the exponential decrease of the ``form factors'' in $V(q,k;{ \rho} )$,
\begin{eqnarray}  
{ {\mathcal Q}\rho
K_1 ({\mathcal Q}\rho )} \stackrel{{\mathcal Q} \rho\rightarrow \infty}
{\rightarrow}\sqrt{\frac{\pi}{2}}\,\sqrt{{\mathcal Q} \rho}\
{ \exp\left[-{\mathcal Q} \rho\right]} 
\, .
\end{eqnarray}
The $\rho$-integral may even be performed analytically, after
inserting the various LSZ amputated FT's into Eq.\,(\ref{Tlsc}).
After continuation to the DIS-regime of Minkowski space, the (effective)
hard scale,
\begin{eqnarray}
{\mathcal Q}\equiv {\rm min}\,\left\{
 Q\equiv \sqrt{ { -q^2}} \, ,\, 
 \sqrt{{ -(q-k_1)^2}}\, , \,\sqrt{ { -(q-k_2)^2}}\right\}\geq 0\,,
\end{eqnarray}
then provides a {\it dynamical IR cutoff for the instanton size}, 
$\rho\lwig {\mathcal O}(1/{\mathcal Q})$.

As a highly non-trivial check of our calculations, 
electromagnetic current conservation,
$q^{\mu} {\mathcal T}_\mu^{{ (I)}\,a\,a_1\ldots a_{n_g}}=0$,
is manifestly satisfied.

Like in perturbative QCD,  
the leading-order $ I$-induced amplitudes are well-behaved as long as we avoid
the collinear singularities, arising when the internal quark
virtualities ${ t\equiv -(q-k_1)^2}$  or 
${ u\equiv-(q-k_2)^2\rightarrow 0}$ vanish in Eq.\,(\ref{vertex})
(c.f. Fig.\,1). Hence high $Q^{2}$ processes of moderate
multiplicity, where {\it the current quark} is produced at a
{\it fixed angle} relative to the photon,  are  reliably calculable 
in instanton perturbation theory. 

Altogether, we may conclude that deep-inelastic scattering is very
well suited for studying manifestations of QCD-instantons!

{\it Example}: We have worked out  explicitly the cross section for
the simplest process~\cite{mrs} for $n_f=1$ without final-state gluons, 
 $ \gamma^{\ast}+{\rm g}\rightarrow 
   \overline{{\rm q}_{L}}+{\rm  q}_{R}\ (\triangle Q_{5}=2)$.

Although being only a {\it small} fraction of the total $I$-induced
contribution, it may be considered as a calculable {\it lower bound}. 
Moreover, it contains all essential features of the dominant multi-gluon
process. The residual dependence on the renormalization scale $\mu_r$ is 
very weak if the 2-loop RG-invariant $I$-density is used~\cite{morretal}. 
In Figs.\,3, we display a comparison with the cross sections for 
the appropriate chirality-conserving process within leading-order
perturbative QCD, $\gamma^{\ast}+{\rm g}\rightarrow 
\overline{{\rm q}_{L}}+{\rm q}_{L}\ (\triangle Q_{5}=0)$.

From the requirement that the average 
instanton size $\langle \rho\rangle$ contributing for a given
virtuality $\mathcal Q$, should be small enough to neglect higher-order 
corrections of $ I$-perturbation theory, a lower
limit on the hard scale $\mathcal{Q}$ can be easily obtained~\cite{mrs},
\begin{equation}
\langle\rho\rangle\,\lwig\, \frac{1}{500 {\rm\ MeV}}< 
 \frac{1}{\Lambda}\Rightarrow \mathcal{Q}\gtrsim {\rm 5} \ {\rm GeV}.
\end{equation}

\begin{figure}
\vspace{-0.4cm}
\begin{center}
\epsfig{file=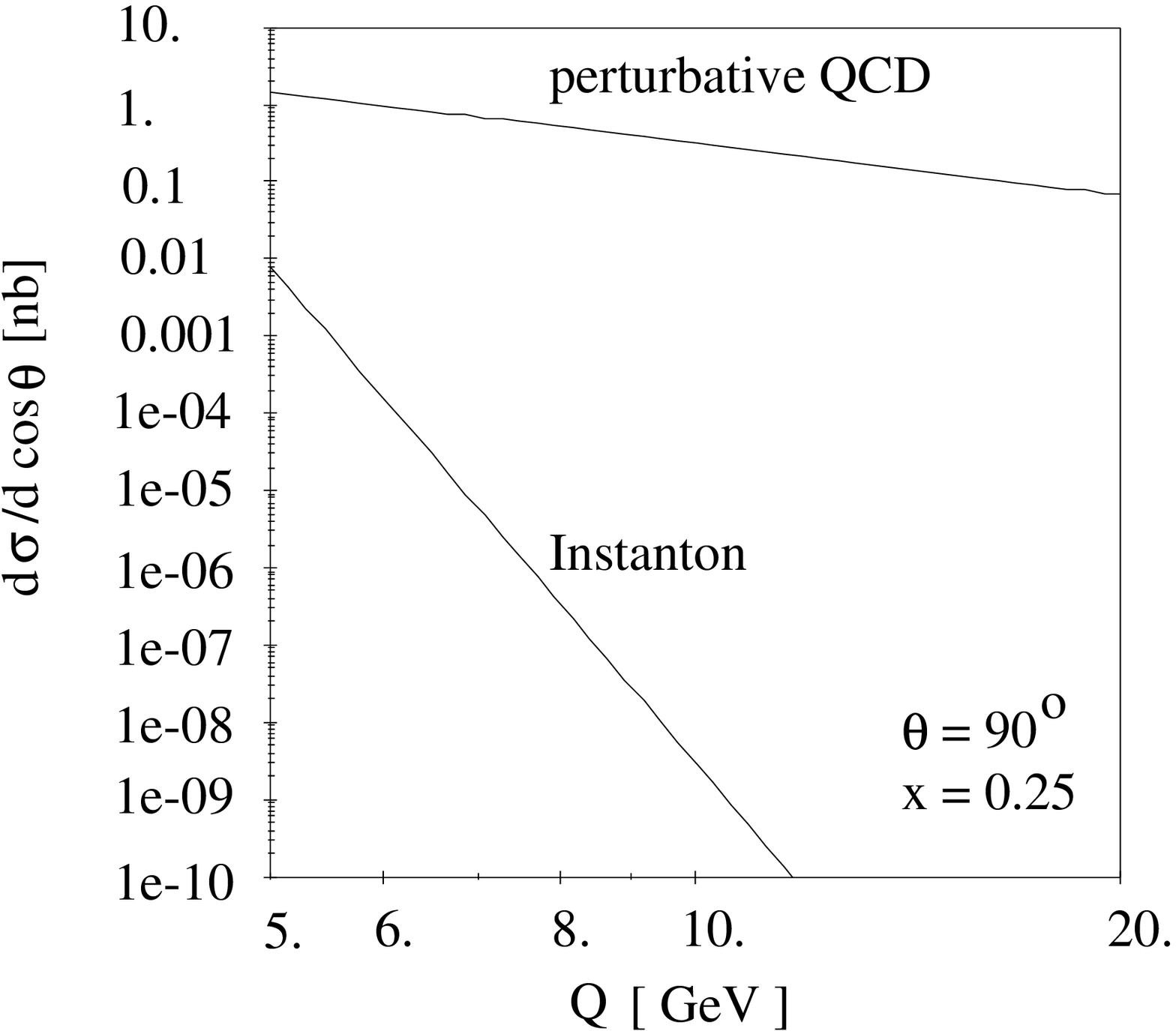,%
width=6cm,height=6cm}
\epsfig{file=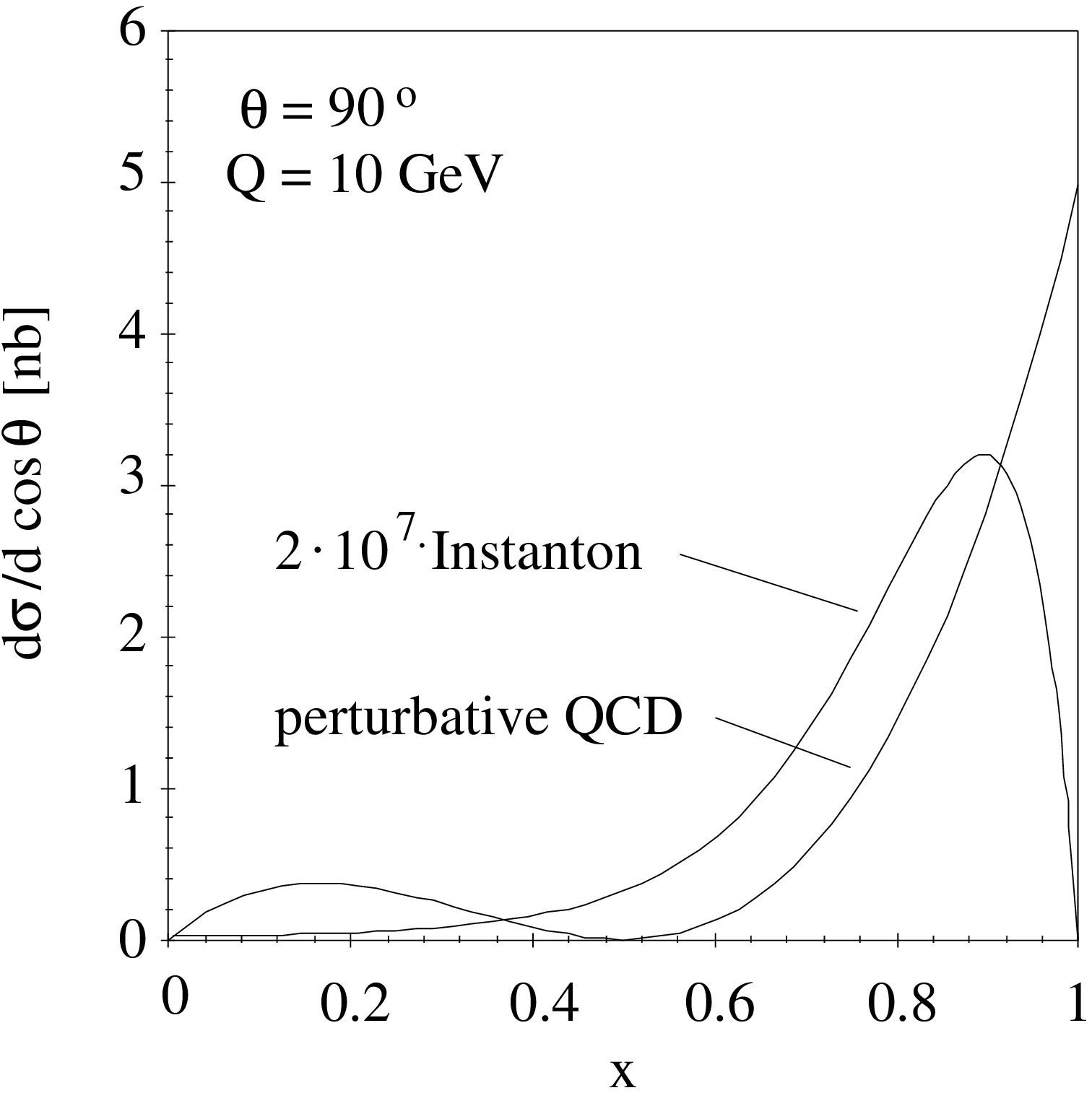,width=6cm}
\caption[dum]{$I$-induced fixed-angle scattering for $\mathcal{Q}\gwig
$5 GeV is well under control. ($e_{q}=2/3,\,\Lambda =234$ MeV,
$\mu_{r}=Q$). }
\end{center}
\end{figure}

\section*{3. Inclusive Approach to  
the Multi-Particle Final State }

A crucial theoretical task is to find the best framework allowing to
make contact with experiment (HERA), without upsetting the validity of
$I$-perturbation theory. Experimentally, the best bet~\cite{rs} is to hunt for 
${ I}$-``footprints'' in the  multi-particle final state rather than
in totally inclusive observables like $F_{\rm 2}(x_{\rm Bj},Q^2)$. Unlike 
the configuration space approach of Ref.~\cite{bb}, our momentum-space
picture allows to keep control over the various ${I}$-approximations also at 
small $x_{\rm Bj}$ through {\it kinematical cuts} on
(reconstructed) {\it final-state momentum variables}\,!

A ``brute-force'' possibility to calculate the desired $I$-induced
cross sections for a {\it multi-parton final state} consists 
simply in squaring our $I$-induced amplitudes from Sect.\,2,
performing the necessary phase-space integrations and summing over 
unobserved partons. For the dominating final states with many gluons
this procedure becomes increasingly tedious and also inaccurate.    

Let us sketch here a second more elegant and presumably also more
accurate approach~\cite{rs3}, which includes an implicit summation
over the {\it exponentiating} leading-order gluon emission as well as
gluonic final-state tree-graph corrections according to the 
$I\overline{I}$-valley method~\cite{y,kr}: 

First of all, it turns out~\cite{rs,rs1,rs2} that 
${\rm d}\sigma^{(I)}\,(\gamma^\ast +g \Rightarrow 2 n_f\,q  + n_g\,g)$
factorizes into a calculable ``splitting''
function associated  with the $\gamma^\ast q q^\ast$-vertex
(c.\,f. Fig.\,1) and cross sections for the $I$-induced {\it
subprocess} $q^\ast g \rightarrow (2 n_f-1)\,q + n_g\,g$, on which we
shall concentrate from now on.
 
Next, we remember that any given final state may be equivalently described 
in terms of $\sigma_{\rm tot}$ and the set of { 1,2,...} parton 
{\it inclusive} cross sections. The latter, in turn,
may be evaluated via the so-called ``Mueller optical
theorems''~\cite{m}, expressing  $n=1,2\ldots$ particle {\it
inclusive} cross sections as  appropriate discontinuities of 
$2+n\rightarrow 2+n$ forward elastic amplitudes in generalization of 
the usual optical theorem (c.\,f. Fig.\,4). Although no rigorous proof
exists, much of the Regge inspired multi-particle phenomenology of the
70's rested on the validity of these  Mueller optical theorems.
\begin{figure}
\begin{center}
\vspace{-0.4cm}
\epsfig{file=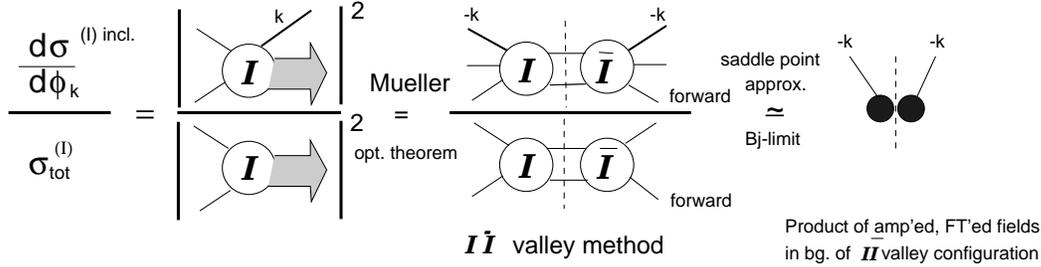,width=14cm}
\caption[dum]{Application of the Mueller optical theorem~\cite{m}, the
$I\overline{I}$-valley method~\cite{y,kr} and  saddle-point integration
over the collective coordinates to estimate the normalized,
$I$-induced  1-parton {\it inclusive} $q^\ast g$ subprocess cross
section; d$\phi_k$ denotes the Lorentz-invariant phase-space element.} 
\end{center}
\end{figure}
The $I$-induced 1,\,2\ldots parton inclusive cross sections can now be
evaluated in complete analogy to existing calculations~\cite{kr} of 
$\sigma^{(I)}_{\rm tot}$ by means of the $I \overline{I}$ valley
method~\cite{y,kr} applied to the $2\rightarrow 2$ forward amplitude in the 
$I\overline{I}$ background. By normalizing the inclusive cross
sections to $\sigma^{(I)}_{\rm tot}$, common, poorly known
pre-exponential factors largely cancel, such that quite stable and accurate
results are obtained (c.\,f. Fig.\,4).

Let us give some examples. From calculating the (normalized) 1-parton
inclusive cross section along these lines, we obtain 
(after phase-space integration) the average gluon and quark
multiplicity  $\langle n_{q+g}\rangle\propto 1/\alpha_s$, as well as  
the {\it isotropy} of $I$-induced parton production  and the transverse 
energy flow vs. pseudo-rapidity $\frac{d\langle E_T\rangle}{d\eta_I}=
\frac{E_I^{\rm tot}}{\langle n\rangle} \frac{1}{\cosh\ \eta_I}$,
both in the c.m.s. of the $I$-subprocess. From the (normalized)
2-parton inclusive cross section, we obtain (after phase-space integration) 
in the Bjorken limit $\langle n^2\rangle - \langle n\rangle^2\approx
0$, implying a {\it Poisson} distribution for the $I$-induced exclusive
n-parton cross sections. Furthermore, we gain information on various
momentum correlations of the produced partons. 

All along, there are most non-trivial consistency conditions 
in form of {\it energy/momentum/charge} sum rules~\cite{tfv}, like e.\,g.
\begin{equation}
\sum_{{\rm q},{\rm g}} \int d\phi_{k}\ k_\mu\ \frac{1}{\sigma_{\rm
tot}^{(I)}}
\frac{d\sigma^{(I)\,{\rm incl.}}}{d\phi_{k}}\ =\  (p+q^\prime )_\mu,
\end{equation}                  
which turn out to be satisfied in the Bjorken limit.

\end{document}